\documentclass[aip, reprint, floatfix]{revtex4-1}

\usepackage{graphicx}
\usepackage{amssymb}
\usepackage{amsmath}
\usepackage{siunitx}

\newcommand{\modified}[1]{#1}

\newcommand{\columbia}{Department of Physics, Columbia University, New York, NY 10027, USA}
\newcommand{\stanford}{Department of Physics, Stanford University, Stanford, CA, 94305-4085, USA}
\newcommand{\slac}{SLAC National Accelerator Laboratory, Menlo Park, CA 94025, USA}
\newcommand{\asusese}{School of Earth and Space Exploration, Arizona State University, Tempe, AZ 85287, USA}

\newcommand{\cardiff}{School of Physics and Astronomy, Cardiff University, Cardiff, Wales CF24 3AA, UK}
\newcommand{\jpl}{NASA, Jet Propulsion Laboratory, Pasadena, CA 91109, USA}

\newcommand{\usc}{Department of Physics and Astronomy, University of Southern California, Los Angeles, CA 90089, USA}

\begin{document}

\title{High quality factor manganese-doped aluminum lumped-element kinetic inductance detectors sensitive to frequencies below 100 GHz}

\author{G.~Jones}
\email{gej2113@columbia.edu}
\affiliation{\columbia}

\author{B.~R.~Johnson}
\affiliation{\columbia}

\author{M.~H.~Abitbol}
\affiliation{\columbia}

\author{P.~A.~R.~Ade}
\affiliation{\cardiff}

\author{S.~Bryan}
\affiliation{\asusese}

\author {H.-M.~Cho}
\affiliation{\slac}

\author{P.~Day}
\affiliation{\jpl}

\author{D.~Flanigan}
\affiliation{\columbia}

\author {K.~D.~Irwin}
\affiliation{\stanford}
\affiliation{\slac}

\author {D.~Li}
\affiliation{\slac}

\author{P.~Mauskopf}
\affiliation{\asusese}

\author{H.~McCarrick}
\affiliation{\columbia}

\author{A.~Miller}
\affiliation{\usc}

\author {Y.~R.~Song}
\affiliation{\stanford}

\author{C.~Tucker}
\affiliation{\cardiff}

\date{\today}


\begin{abstract}
Aluminum lumped-element kinetic inductance detectors (LEKIDs) sensitive to millimeter-wave photons have been shown to exhibit high quality factors, making them highly sensitive and multiplexable.
The superconducting gap of aluminum limits aluminum LEKIDs to photon frequencies above 100~GHz.
Manganese-doped aluminum (Al-Mn) has a tunable critical temperature and could therefore be an attractive material for LEKIDs sensitive to frequencies below 100~GHz if the internal quality factor remains sufficiently high when manganese is added to the film.
To investigate, we measured some of the key properties of Al-Mn LEKIDs.
A prototype eight-element LEKID array was fabricated using a \SI{40}{nm} thick film of Al-Mn deposited on a \SI{500}{\micro m} thick high-resistivity, float-zone silicon substrate.
The manganese content was 900~ppm, the measured $T_c = 694\pm1$\SI{}{mK}, and the resonance frequencies were near 150~MHz.
Using measurements of the forward scattering parameter $S_{21}$ at various bath temperatures between 65 and 250~mK, we determined that the Al-Mn LEKIDs we fabricated have internal quality factors greater than $2 \times 10^5$, which is high enough for millimeter-wave astrophysical observations.
In the dark conditions under which these devices were measured, the fractional frequency noise spectrum shows a shallow slope that depends on bath temperature and probe tone amplitude, which could be two-level system noise.
The anticipated white photon noise should dominate this level of low-frequency noise when the detectors are illuminated with millimeter-waves in future measurements. 
The LEKIDs responded to light pulses from a 1550~nm light-emitting diode, and we used these light pulses to determine that the quasiparticle lifetime is 60~$\mu$s. 
\end{abstract}


\maketitle



\begin{figure}[t]
\includegraphics[width=\columnwidth]{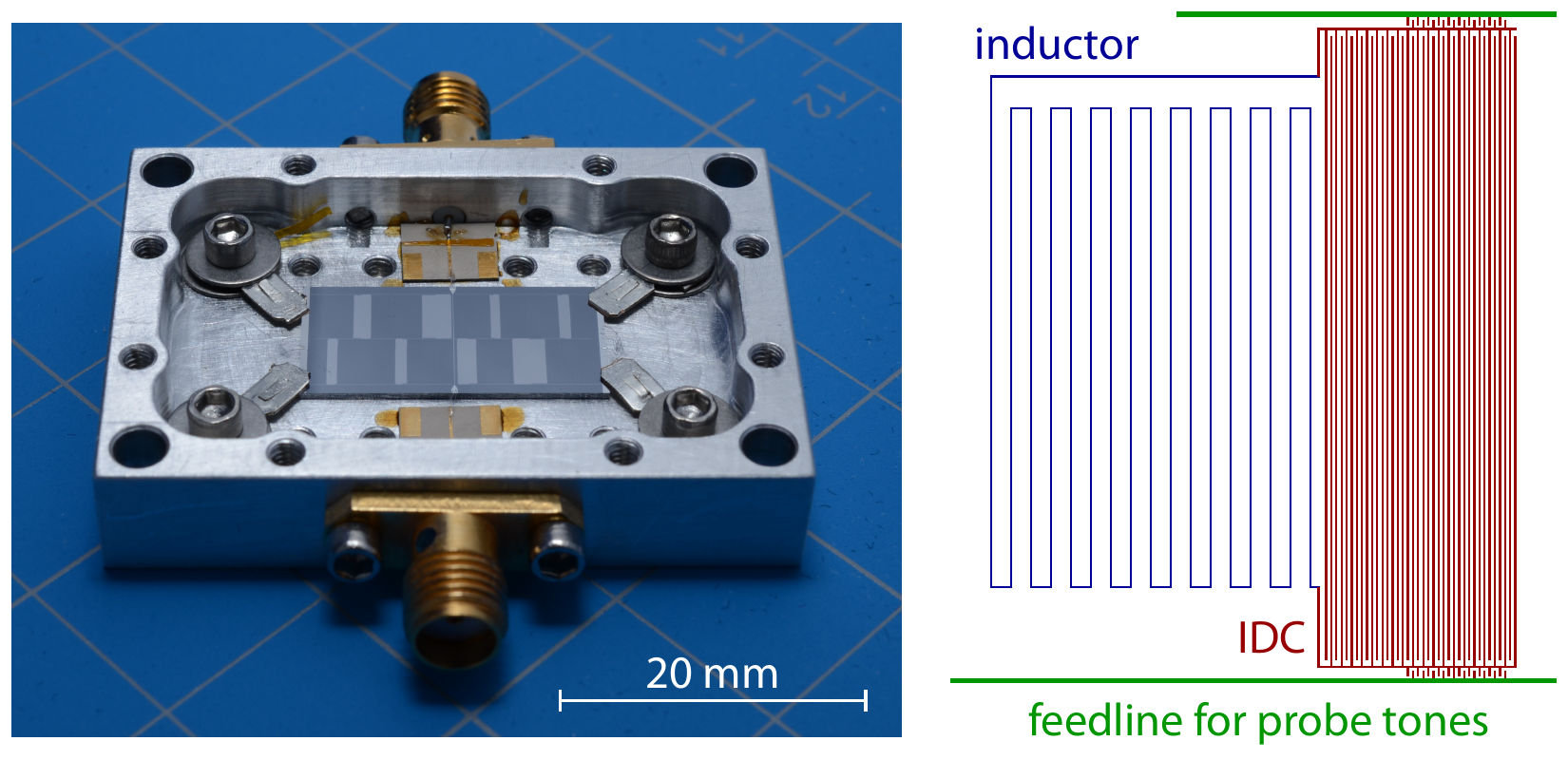}
\caption{
\textbf{Left:} A photograph of the detector module tested in this study.
The package lid is removed so the LEKID array is visible.
Metal clips are used in the corners to hold the LEKID array in place.
\textbf{Right:} A schematic drawing of one LEKID in the array.
The signal from the feedline is capacitively coupled to the resonator.
}
\label{fig:array}
\end{figure}

Lumped-element kinetic inductance detectors (LEKIDs) are superconducting thin-film resonators that are designed to detect photons energetic enough to break Cooper pairs\cite{Doyle2010,Day2003}.
Each resonator consists of an interdigitated capacitor (IDC) and a meandered inductor, which absorbs the incident radiation directly (see Figure~\ref{fig:array}).
LEKIDs are sensitive to photons with frequencies greater than $\nu_c = 2\Delta/h \approx \mbox{74~GHz} \times T_c/(\mbox{1~K})$, where $\Delta \approx 1.76 k_B T_c$ is the superconducting gap, $h$ is Planck's constant, $k_B$ is Boltzmann's constant, and $T_c$ is the critical temperature of the superconductor.
Thin-film aluminum LEKIDs are being developed for cosmic microwave background (CMB) studies near 150~GHz\cite{mccarrick_2016,McCarrick2014}.
Thin-film aluminum works well for this spectral band because the $T_c \sim 1.4$~K, and the associated low-frequency cutoff is approximately 100~GHz.
However, future CMB observations, such as CMB Stage-4, will observe at a range of frequencies including frequencies below 100~GHz\cite{cmb-s4_2016}.
Thus materials with a tunable $T_c$ are particularly attractive because the LEKIDs can be re-optimized for the lower photon frequencies of interest.
Sub-stoichiometric titanium nitride (TiN$_x$) and TiN/Ti/TiN trilayers are already being considered for these lower-frequency devices\cite{leduc_2010,Lowitz_2016}.
An alternate material is manganese-doped aluminum (Al-Mn).
The $T_c$ of thin-film aluminum can be reduced in a controllable way by adding a small amount of manganese\cite{deiker_2004,barends_2009,Young2004,Boato1966}.
High-purity Al-Mn sputtering targets can be used to deposit extremely uniform films, and these targets can be purchased with a specified manganese content making the film $T_c$ very predictable and reproducible.
Because of these attractive properties, Al-Mn is now commonly used to fabricate transition edge sensor bolometers\cite{li_2016,Schmidt2011}.
The high sensitivity and dense multiplexing factors of LEKIDs rely on the high quality factors ($Q$) that can be achieved when the superconducting films are operated below $\sim T_c/5$.
\modified{Because manganese can be a magnetic material, one concern is that it could adversely affect the $Q$ of the detectors\cite{flanigan_2016}.
However, it has been demonstrated that manganese loses its magnetic character in lightly doped Al-Mn, and the resulting films have been shown to retain a sharp density of states, with only a slight broadening of the superconducting gap, as expected for non-magnetic impurities\cite{Kaiser1970,ONeil2008,ONeil2010}.
We therefore expect LEKIDs made from Al-Mn films to have the same high quality factors seen in pure aluminum devices.}
\modified{To test this hypothesis,}
we built and tested an array of eight Al-Mn LEKIDs to characterize their quality factors, and our results are presented in this paper.
%
%


The LEKID array was fabricated using a \SI{40}{nm} thick film of Al-Mn sputter deposited onto a \SI{500}{\micro m} thick high-resistivity, float-zone silicon substrate that was pre-cleaned with HF to remove native oxides.
The atomic ratio of manganese to aluminum was 900~ppm.
For this demonstration, the detectors were mounted in a light-tight aluminum package and the seams were sealed with copper tape.
Figure~\ref{fig:array} shows the LEKID array in the package.
The test array consists of eight resonators patterned on a \SI{20}{mm}$\times$\SI{10}{mm} chip.
The designed resonance frequencies were in the range 100--200~MHz.
The integrated module was cooled using an adiabatic demagnetization refrigerator (ADR) backed by a helium pulse tube cooler\cite{McCarrick2014}.
A $\mu$-metal shield was placed around the outside of the cryostat to suppress the effects of ambient magnetic fields\cite{flanigan_2016}.

\begin{figure}[h]
\includegraphics[width=\columnwidth]{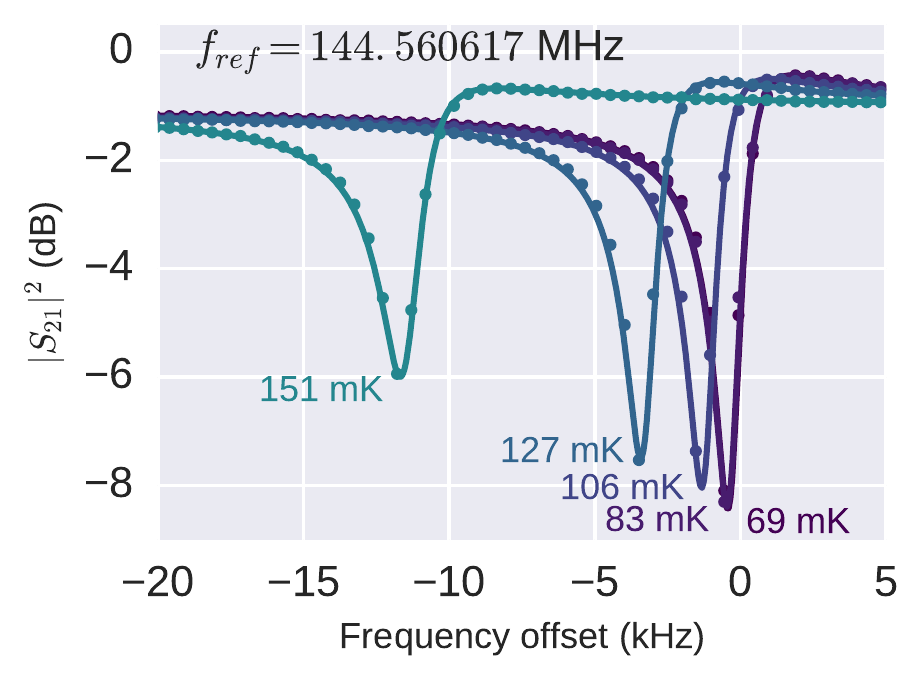}
\caption{
Measured microwave transmission ($S_{21}$) sweeps at a range of module temperatures.
The points are measured data and the lines show the best fit resonator model.
}
\label{fig:s21_vs_f}
\end{figure}

At a given bath temperature, for each resonator, we collected two data sets.
First, we measured the forward scattering parameter $S_{21}$ by sweeping the readout tone frequency across the resonance.
This data set provided the resonance frequency.
Second, we set the readout tone to the measured resonance frequency and collected time-ordered data for \SI{120}{s}.
We identified six of the eight resonators in the array, all of which showed similar performance.
The data presented in this paper is all from a single resonator with a resonance frequency near \SI{144.56}{MHz}.
The $S_{21}$ sweeps for five representative module temperatures are shown in Figure~\ref{fig:s21_vs_f}.
The measurements presented here were recorded using readout tone powers of approximately \SI{-114}{dBm} to \SI{-111}{dBm} on the feedline.
This is several dB below the power at which the resonator bifurcates, which was measured to be approximately \SI{-104}{dBm}.
The corresponding non-linear inductance scaling energy\cite{Swenson2013} $E_{\star}$ is within a factor of two of the condensation energy for the inductor $E_c=N_0 \Delta ^2 V_L /2$, assuming the same single-spin density of states at the Fermi level $N_0=\SI{1.72e10}{eV^{-1} \micro m ^{-3}}$ as for pure aluminum. The inductor volume $V_L=\SI{2800}{\micro m^3}$.


We determined the resonance frequency $f_r$, the total quality factor $Q$, and the coupling quality factor $Q_c$ by fitting a model including a resonance term of the form
\begin{align}
S_{21} = 1 - \frac{Q}{Q_c} \frac{1}{1 + 2 j Q x},
\end{align}
as well as multiplicative terms to account for cable delay and gain slope, to the measured $S_{21}$ sweeps\cite{McCarrick2014,Khalil2012}.
We define the fractional frequency shift $x=1-f_r/f_{ref}$, where $f_{ref}$ is a reference frequency, typically taken to be the maximum observed resonance frequency.
The coupling quality factor of the resonator studied here is approximately $150 \times 10^3$ as designed.
From $Q$ and $Q_c$ we calculate the inverse internal (unloaded) quality factor, which can be thought of as the loss in the resonator, using the expression $1/Q_i = 1/Q - 1/Q_c$.  
We also use this model to transform the $S_{21}$ time series data into fluctuations in $x$ and then use standard spectral analysis to compute the power spectral density of the fractional frequency fluctuation, $S_{xx}$.

\begin{figure}[t]
\includegraphics[width=\columnwidth]{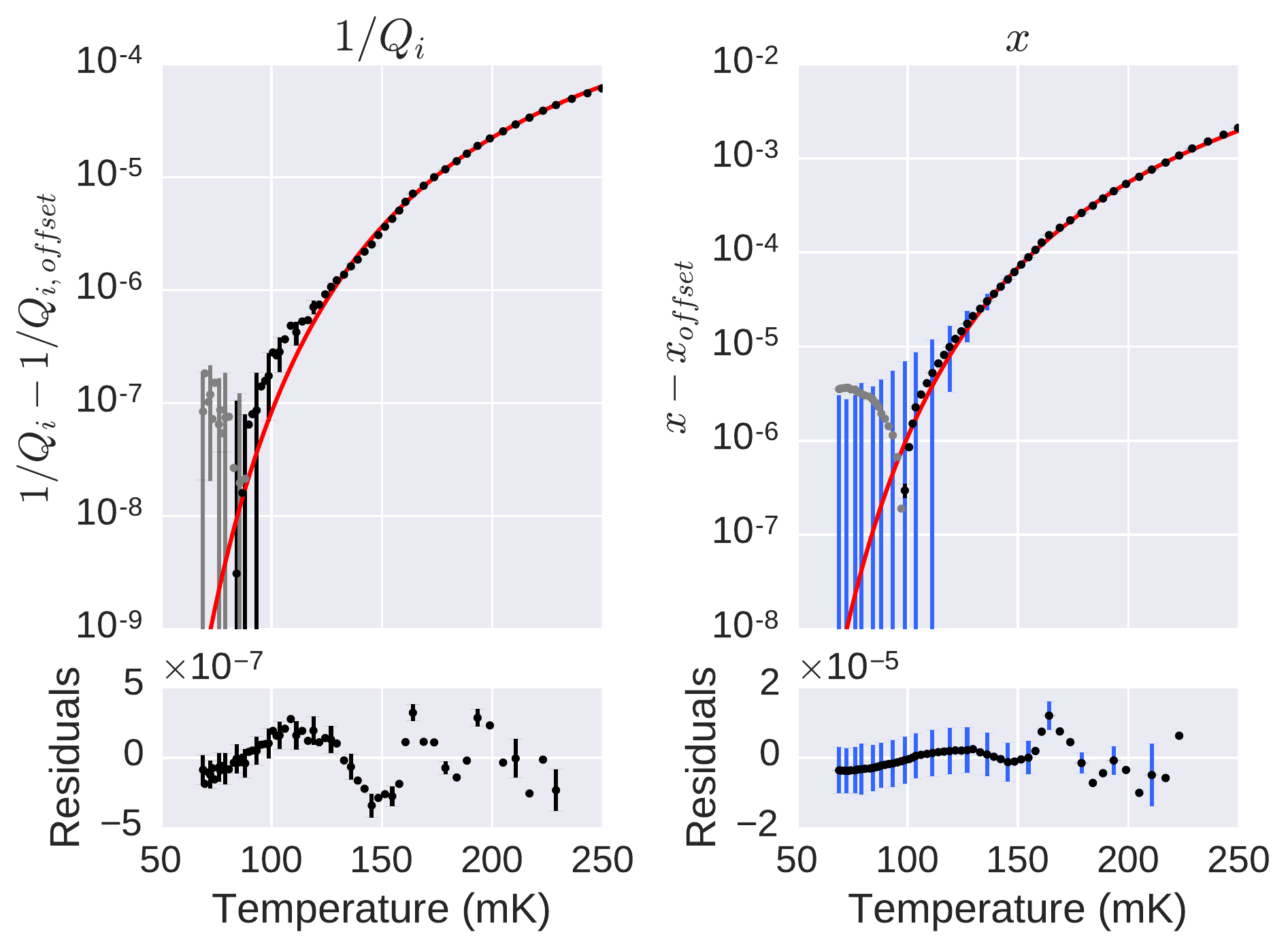}
\caption{
\modified{
Inverse quality factor (left) and fractional frequency shift (right) as a function of module temperature.
The measured data points and error bars are shown in red in the top panels.
For clarity, every third error bar is shown.
The best-fit Mattis-Bardeen model is shown in red.
The residuals are shown in the bottom panels.
The visible (blue) error bars in the $x$ plots have been expanded by a factor of 129 for the joint fit as described in the text.
The gray points in the two upper panels have negative values (i.e. the measured value of $x$ or $1/Q_i$ is less than the best fit $x_{offset}$ or $1/Q_{i,offset}$, respectively), and so the absolute value is plotted.}
}
\label{fig:iQi_x_vs_T}
\end{figure}

The inverse internal quality factor and $x$ are plotted as a function of module temperature in Figure~\ref{fig:iQi_x_vs_T}.
\modified{Previous experiments have demonstrated that aluminum films with similar levels of manganese doping retain sharp energy gaps in their density of states because the manganese impurities lose their magnetic characteristics\cite{Kaiser1970,ONeil2008,ONeil2010}}.
\modified{We therefore fit the data presented in Figure~\ref{fig:iQi_x_vs_T}}
 to equations derived from the Mattis-Bardeen integrals\cite{Zmuidzinas2012}\modified{, which are commonly used to model aluminum LEKIDs\cite{Gao2008}}:
\begin{align}
1/Q_i(T) =& \frac{4\alpha_q}{\pi}\exp\left(\frac{-\Delta}{k_B T}\right)\sinh{(\xi)}\mathrm{K_0}(\xi) \nonumber \\ +&~1/Q_{\mathrm{TLS}}(T,F_{\mathrm{TLS}} \delta_0) + 1/Q_{\mathrm{i,offset}} \\
x(T) =& \frac{\alpha_x}{2\Delta}\sqrt{2\pi k_B T \Delta} \, \exp\left(\frac{-\Delta}{k_B T}\right) \nonumber \\ \times& \left(1+\sqrt{2\pi k_B T \Delta} \, \exp{(-\xi)} \, \mathrm{I_0}(\xi)\right) \nonumber \\ -&~x_{\mathrm{TLS}}(T,F_{\mathrm{TLS}} \delta_0) + x_{\mathrm{offset}}.
\end{align}
Here $\xi=hf_r/(2 k_B T)$, where $f_r$ is the resonance frequency, and $I_0$ and $K_0$ are modified Bessel functions of the first and second kind.
The TLS contribution as a function of temperature is included as $Q_\mathrm{TLS}(T,F_{\mathrm{TLS}} \delta_0)$ and $x_\mathrm{TLS}(T,F_{\mathrm{TLS}} \delta_0)$, which are defined in the literature\cite{Zmuidzinas2012}.
We fit for the value of the product of the TLS filling factor $F_{\mathrm{TLS}}$ and low-temperature loss tangent $\delta_0$ as a single parameter.
We set $\Delta=1.76k_B T_c$ and fit for $T_c$.
For the joint fit to succeed, we find it necessary to allow different scale factors $\alpha_q$ and $\alpha_x$ for the $Q_i$ and $x$ response, respectively. 
The fits show no evidence for TLS in the $S_{21}$ sweeps and imply an upper limit on $F_{\mathrm{TLS}} \delta_0 < 3 \times 10^{-7}$.
The $1/Q_{\mathrm{offset}}$ and $x_{\mathrm{offset}}$ terms are nuisance parameters that absorb any bias from not knowing the true zero-temperature behavior of the resonator. 
We also included a nuisance parameter to scale the uncertainties in the $x$ measurements to prevent the joint fit from being completely dominated by the $x$ data. 
This resulted in scaling the $x$ uncertainties by a factor of 129.
The uncertainties before and after scaling, are shown in Figure~\ref{fig:iQi_x_vs_T}.

We find that the joint fit implies $T_c=694\pm1$\SI{}{mK}, assuming $\Delta = 1.76 k_B T_c$.
This is somewhat lower than the value of $730\pm10$~mK observed by monitoring the feedline transmission during warm up, which could imply that $\Delta = 1.66 k_B T_c$ for this sample.
A similar reduction in the effective gap implied by fitting the Mattis-Bardeen model to resonance frequency versus temperature data has been observed in TiN resonators, and was attributed to broadening of the superconducting gap\cite{Bueno2014}.
\modified{Other studies in the literature show that a slight broadening of the gap in Al-Mn is expected, and has been modeled using a so-called Dynes parameter\cite{Kaiser1970,ONeil2008,ONeil2010}.
Because it is not expected to significantly impact the optical sensitivity of the resulting LEKIDs, investigating the details of this gap broadening effect is beyond the scope of this initial study.
However, we note that gap broadening may also explain some of the structure in the residuals shown in Figure~\ref{fig:iQi_x_vs_T}.
}
The scale factors $\alpha_x=0.168$ and $\alpha_q=0.106$ are reasonable when interpreted as kinetic inductance fractions, and similar differences in the loss and fractional frequency responses have been observed in other aluminum resonators\cite{Budoyo2016}.
Non-dissipative quasiparticle states that affect $x$ but not $Q_i$ have also been invoked to explain other MKID data\cite{Gao2012}.

%
%
%
%

\begin{figure}[h]
\includegraphics[width=\columnwidth]{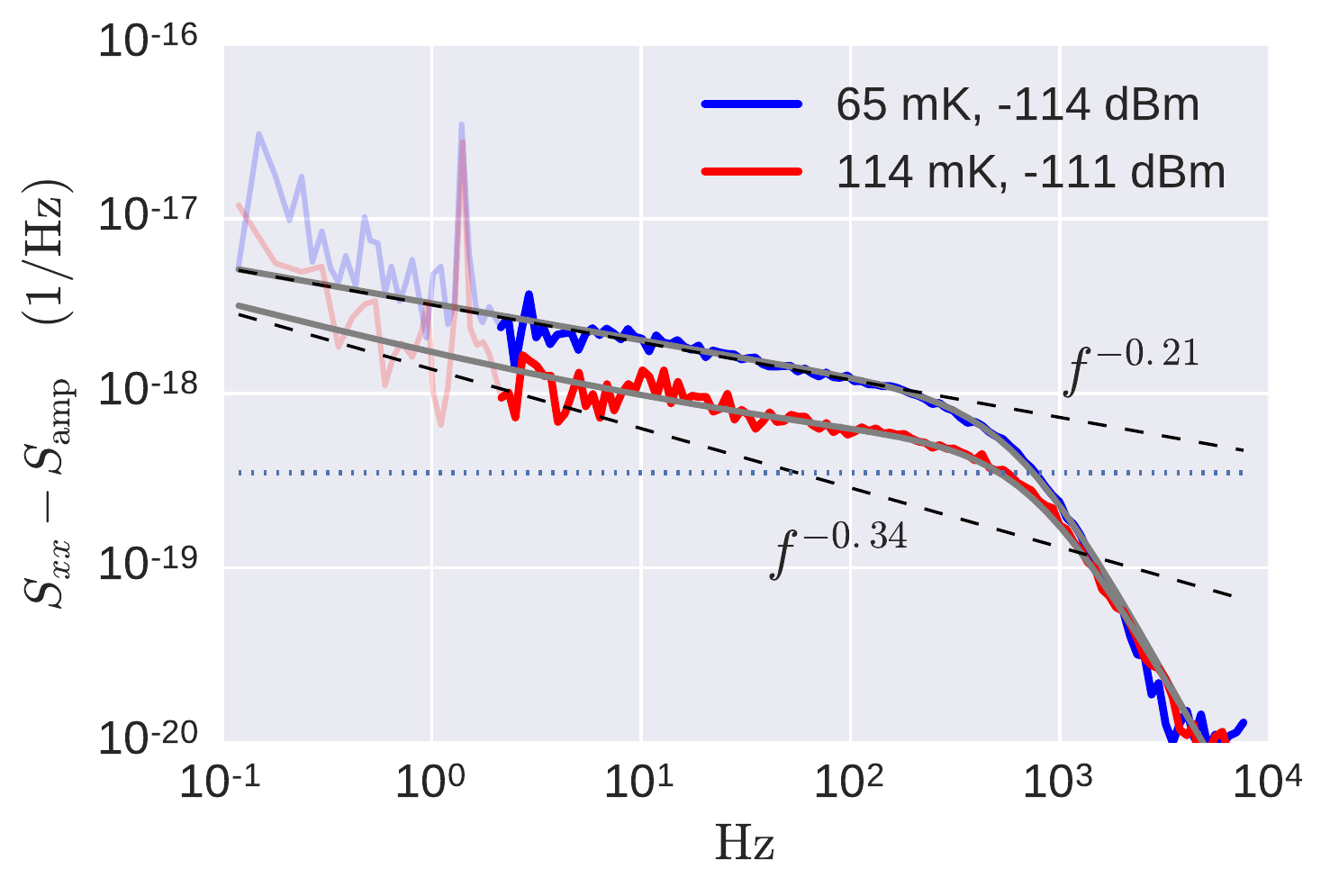}
\caption{
The spectral density of the fractional frequency time series data at module temperatures \SI{65}{mK} and \SI{114}{mK}, as well as probe tone powers of $-114$ and $-111$~dBm at the feedline.
To ease comparison between the different test conditions the noise floor set by the low noise amplifier ($S_{amp}$) was modeled and subtracted from the data.
The data have been binned in logarithmic frequency intervals for clarity. The dark portion of each curve shows the data used in the fits. 
The best fit models are shown by the gray lines.
The black dashed lines show the red noise components of each fit.
The blue dotted line shows the white noise level in the detector band ($S_{w}$) for the 114~mK data.
The spur at \SI{1.4}{Hz} is due to the pulse tube cooler.
The steeper noise below \SI{1}{Hz} is largely correlated between detectors and is suspected to be caused by module temperature fluctuations and microphonics.
}
\label{fig:Sxx_vs_f}
\end{figure}

Noise spectra from a single resonator are shown in Figure~\ref{fig:Sxx_vs_f} at module temperatures \SI{65}{mK} and \SI{114}{mK}, and with probe tone powers differing by \SI{3}{dB}.
At \SI{65}{mK} the noise is entirely dominated by a shallow power law $\propto \nu ^ {-0.2}$.
At \SI{114}{mK} there is evidence for a white noise component, but a slightly steeper power law dominates below $\sim$\SI{20}{Hz}. 
The noise spectra were fit to
\begin{equation}
S_{xx}(f) = \left(S_{w}+A_1 f^{\beta}\right)\left(1+\frac{f^2}{f_{3\,dB}^2}\right)^{-1}+S_{amp},
\end{equation}
where $A_1$ is the red noise power at \SI{1}{Hz}, $S_{w}$ is the level of the white component of the device noise, $\nu_{3\,dB}$ is the \SI{3}{dB} roll-off frequency, and $S_{amp}$ is the LNA contribution to the fractional frequency noise.
The roll-off frequencies are generally comparable to the resonator ring-down bandwidth $f_r/(2Q)$.

To further investigate the resonator time constant, we performed an additional experiment where we modified the setup by coupling light from a 1550~nm wavelength LED to illuminate the resonators via an optical fiber that passed through a small hole in the package.
We then observed the resonator response at a higher order resonant mode around 1000~MHz.
This resonance had $Q\sim 3 \times 10^4$ and $Q_i \sim 6 \times 10^5$, thus the resonator ring-down bandwidth was around \SI{16}{kHz}.
Using pulses of light from the LED, this wider bandwidth allowed us to measure a time constant of $\sim \SI{60}{\micro s}$ at 65~mK.
This is consistent with other measurements of aluminum LEKIDs under similar conditions\cite{Jones2015}.


The Al-Mn LEKIDs we fabricated have high internal quality factors, limited at the lowest temperatures ($65$~mK) to approximately $2 \times 10^5$, which is high enough for millimeter-wave astrophysical observations.
We also observed $Q_i$ as high as $6 \times 10^5$ at resonance frequencies around 1~GHz, which suggests that this material will also be useful for traditional MKIDs based on transmission line resonators with GHz resonance frequencies, including the multichroic millimeter wavelength detectors currently being designed and tested by our collaboration\cite{Johnson2016}.
The limiting loss mechanism is currently unknown, but could result from residual magnetic fields or an unanticipated quasiparticle population excited by some feature of our experimental setup.
In the dark conditions under which these devices were measured, the fractional frequency noise spectrum shows a shallow slope, which could be TLS noise.
It is likely this noise level will be sub-dominant to white photon noise under optical loading conditions that are typical for for ground-based instruments.
As a next step, we will verify this by illuminating these LEKIDs and studying their noise properties in more detail.
An added advantage of low-$T_c$ films is that photons with frequencies much greater than $\nu_c$ will be energetic enough to break multiple Cooper pairs in the sensing element, which should further suppress the detector noise relative to the photon noise.
We will investigate this hypothesis as well with future measurements.



\modified{We thank the anonymous reviewers for their very thoughtful comments that significantly improved this paper.}
This project is supported by grants from the National Science Foundation to B.~R.~J., K.~I., and P.~M. (Awards \#1509211, \#1509078, and \#1506074) and by a grant from the Research Initiatives for Science and Engineering program at Columbia University to B.~R.~J.
H.~M. is supported by a NASA Earth and Space Sciences Fellowship.
We acknowledge the Stanford Nanofabrication Facility, where the devices were fabricated.
We thank the Xilinx University Program for their donation of the FPGA hardware and software tools that were used in the readout system. 


\bibliography{references}

\end{document}